# Verification of Operational Numerical Weather Prediction Model Forecasts of Precipitation Using Satellite Rainfall Estimates over Africa

Yan Wang[1,‡], Moussa Gueye[1,8,‡], Steven J. Greybush[1,3,§], Helen Greatrex[2,3,§], Andrew J. Whalen[4,7], Paddy Ssentongo[4,5], Fuqing Zhang[1,†], Gregory S. Jenkins[1], and Steven J. Schiff[3,4,6,§,*]

January 6, 2022

[1]Department of Meteorology and Atmospheric Science, The Pennsylvania State University, University Park, PA, USA.

[2]Department of Geography and Department of Statistics, The Pennsylvania State University, University Park, PA, USA

[3]Institute for Computational and Data Sciences, The Pennsylvania State University, University Park, PA, USA

[4]Center for Neural Engineering and Center for Infectious Disease Dynamics, The Pennsylvania State University, University Park, PA, USA.

[5]Public Health Science, The Pennsylvania State University, University Park, PA, USA.

[6]Departments of Neurosurgery, Engineering Science and Mechanics, and Physics, The Pennsylvania State University, University Park, PA, USA

[7]Department of Neurosurgery Massachusetts General Hospital, Harvard University, MA, USA

[8]Université du Sine Saloum El Hadji Ibrahima Niass (USSEIN), Kaolack, Senegal

‡, § Contributed Equally

† Deceased

*To whom correspondence should be addressed

**Key Points:**

1. Skill metrics (e.g. bias, RMSE, correlation coefficient, and dichotomous statistics) of ECMWF and GFS forecasts against satellite rainfall estimates show spatial and seasonal variations over Africa.

2. The Numerical Weather Model (NWM) forecasts have better skill at predicting rainfall using a weekly aggregation time than a daily aggregation time.

3. Positive skill of NWM forecasts over a number of regions and seasons indicate promise to inform agricultural and public health applications, although there is room for improvement in forecast capabilities.

**Abstract**

Rainfall is an important variable to be able to monitor and forecast across Africa, due to its impact on agriculture, food security, climate related diseases and public health. Numerical Weather Models (NWM) are an important component of this work, due to their complete spatial coverage, high resolution, and ability to forecast into the future. In this study, we seek to evaluate the spatiotemporal skill of short-term rainfall forecasts of rainfall across Africa. Specifically, the European Centre for Medium-Range Weather Forecasts (ECMWF) and the National Centers for Environmental Prediction-Global Forecast System (NCEP-GFS) forecast models are verified by Rainfall Estimates 2.0 (RFE2) and African Rainfall Climatology Version 2 (ARC2), which are fused products of satellite and in-situ observations and are commonly used in analysis of African rainfall. We found that the model rainfall forecasts show good consistency with the satellite rainfall observations in spatial distribution over Africa on the seasonal timescale. Evaluation metrics show high spatial and seasonal variations over the African continent, including a strong link to the location of the inter-tropical convergence zone (ITCZ) and topographically enhanced precipitation. The rainfall forecasts at one week aggregation time are improved against daily forecasts.

**Plain Language Summary**

Sub-Saharan Africa is extremely vulnerable to rainfall hazards, whether through rainfall related activities (such as rain-fed farming), through the direct impacts of severe weather, or through more indirect impacts such as the impact of rainfall on public health. Therefore the ability to both monitor and forecast rainfall across the region is exceedingly important. This study explores the accuracy of two numerical weather model rainfall forecasts, which are compared against two satellite rainfall estimates over Africa at daily, weekly, and seasonal scales.

Graphical Table of Contents

**Verification of Operational Numerical Weather Prediction Model Forecasts of Precipitation Using Satellite Rainfall Estimates over Africa**

Yan Wang, Moussa Gueye, Steven J. Greybush, Helen Greatrex, Andrew J. Whalen, Paddy Ssentongo, Fuqing Zhang, Gregory S. Jenkins, and Steven J. Schiff*

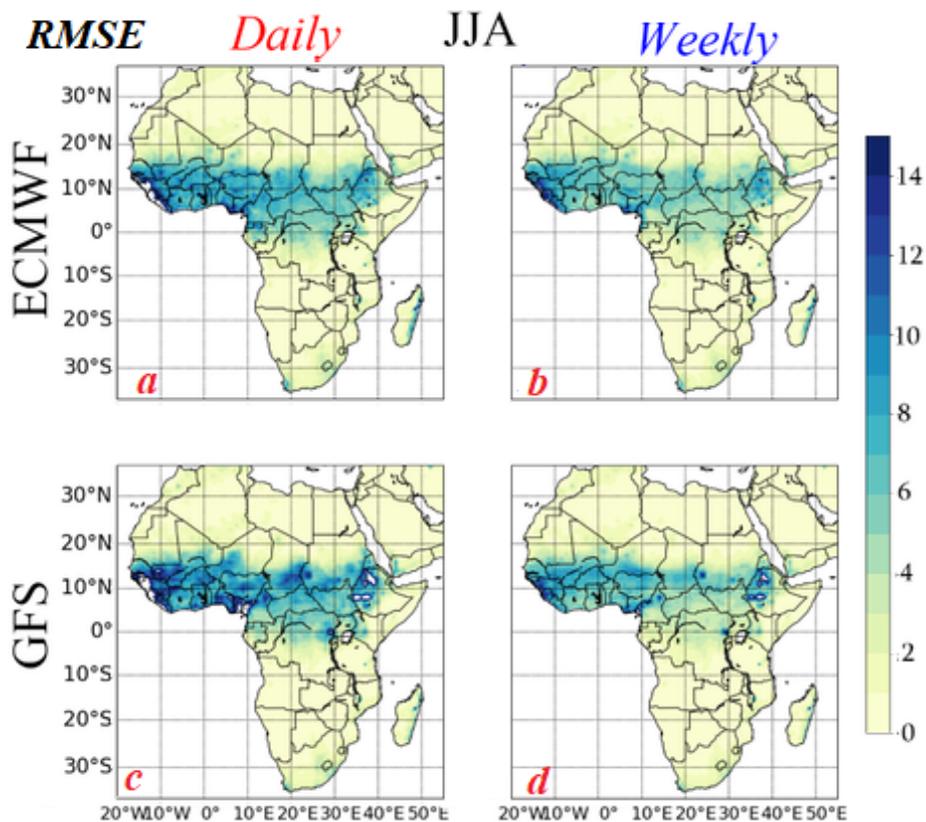

This manuscript evaluates the performance of Numerical Weather Model rainfall forecasts against satellite rainfall estimates over Africa using several metrics. For example, root mean square errors, RMSE (mm day$^{-1}$), vary spatially across the African continent, and vary by season, prediction model (panels a,b versus c,d) and whether the rainfall amounts are aggregated daily (a,c) or weekly (b,d).

# 1. Introduction

## 1.1 Motivation

Numerical weather prediction (NWP) provides tremendous societal benefit, from the advanced warning of weather hazards to planning for economic activities such as agriculture (Alley et al., 2019). Improvements of in situ and remotely sensed observations, numerical model simulations, including their resolution and representation of physical processes, and the data assimilation techniques used to ingest these observations have led to the improved practical predictability of many weather phenomena (Houtekamer and Zhang, 2016). However, the availability of sufficient observations and regional models to forecast rainfall is limited in the African region, proving a challenge for numerical weather prediction.

Operational numerical weather models (NWM) are used to produce short-term and seasonal forecasts across the globe. In many countries, global NWP centers are well connected with their National Meteorological Agencies or Forecasting Offices, where forecasters are well trained and make best use of NWP products. The variations in skill of NWM across African geographies and climatologies is well studied(Gebrechorkos et al., 2018; Kniffka et al., 2020; Linden et al., 2020; Maidment et al., 2013; Maidment et al., 2017; Ogutu et al., 2017; Taraphdar et al., 2016; Tompkins and Feudale, 2010 amongst others). However, a large proportion of research efforts have focused on seasonal or climate timescales, or restricted analysis to specific regions of Africa, or individual weather phenomena or events. Such studies are essential and provide high accuracy but are less helpful in understanding NWM performances comprehensively across the African continent as a whole. In addition, few studies have targeted the NWM performance of daily or weekly forecasts over a long time range, despite those having large societal impact (next Section).

One of the reasons underpinning fewer large-scale assessments of NWM accuracy is that studies are often limited by the availability of ground-based observing networks for in situ verification of weather conditions, especially rainfall. While rain gauges provide relatively accurate and trusted measurements of precipitation at single points, their coverage is lacunar over many African regions. Remotely sensed rainfall products have been available since the early 1980s to fill this gap, supplementing a recent drop in available rain-gauge information (Washington and Downing, 1999).

## 1.2 Implications for public health

Along with benefits for sectors such as agricultural monitoring and disaster risk management, the ability to better forecast short-term weather variables has especially important implications for public health (Heaney et al., 2016; Thomson et al., 2018; Thomson and Mason, 2019). Rainfall is vital to the dynamics of many infectious diseases. It also has a strong relationship with the seasonal variations of diseases in Africa such as seasonal Neisseria meningitis (Agier et al., 2017), Malaria (Pascual et al., 2008), Cholera (Koelle et al., 2005), or Hemorrhagic Fever (Bi et al., 1998).

The effect of rainfall on disease goes beyond seasonality, as short term rainfall events can have a disproportionately large impact on disease outcomes. For example, Lemaitre et al (2019) found that strong precipitation events modify the intra-seasonal double peak of epidemic cholera. Hand, foot and mouth disease (HFMD) also shows a complex association with weekly temperature and precipitation, with Hii et al. (2011) finding that in the Asia Pacific region weekly cumulative rainfall below 75 mm has a positive effect on transmission risk, while rain above 75 mm has a negative effect. When studying dengue, Hooshyar et al. (2020) showed that smaller rainfall peaks over March-April-May (MAM) could lead to larger case peaks over June-July-August (JJA). Equally, it has been found that heavy rainfall may quench and reduce dengue transmission, but dry spells in some settings could increase water storage in the environment, making it more suitable for breeding and propagation of the arthropod vectors that carry the disease (Fouque et al., 2006; Hii et al., 2012). Finally, and with relevance to the larger study this work is part of, bacterial infections leading to postinfectious hydrocephalus in East Africa appears to have a seasonal variation with rainfall (Paulson et al., 2020; Schiff et al., 2012), as does the bacterial disease melioidosis in Southeast Asia and Northern Australia (Wiersinga et al., 2018).

We suggest that high quality forecasts of rainfall and other climate variables is a necessary and required component of what we term predictive personalized public health (P3H), where point-of-care decision-making can be improved through geospatial mapping that takes into account the functional relationships of the dynamics of infectious disease and environmental factors. To do this, it is increasingly important to assess the short-term rainfall characteristics of available NWM models.

## 1.3 Aim of this work

The aim of this study is to assess the performance in characterizing variations in precipitation of two operational numerical weather models, ECMWF and NCEP-GFS, across the continent of Africa. These were assessed at a continental scale for the years 2016-2018; a time frame selected both because these were years where contemporary NWP forecasts driven by modern observing systems were available across all products, and due to the relevance of 2016-2018 for a parallel public health predictive effort (Paulson et al 2020).

Satellite rainfall estimates were selected as validation datasets due to the uneven spatiotemporal distribution of ground-based weather networks across Africa. Specifically, NOAA ARC2 and RFE2 were chosen because of their common operational usage in African rainfall monitoring. As satellite rainfall products are themselves imperfect estimates, multiple validation datasets were selected in order to reduce dependence on individual product characteristics.

In this paper, Section 2 describes the NWM and satellite rainfall observation datasets, and Section 3 describes the methodology for validation. Because the importance of rainfall in public health applications goes beyond daily rainfall forecasts, this study assesses NWM forecast performance at daily, weekly and seasonal scales in Section 4. Finally, in section 5 we consider the implications for using these forecasts as inputs to public health prediction systems in the future.

## 2. Data

### 2.1 Observational dataset, NOAA RFE 2.0 and ARC 2.0

Satellite rainfall datasets are homogeneous and gridded, with a high resolution and little missing data, proving advantageous for assessing NWM forecast accuracy. However there are multiple algorithms available, which have been widely shown to have varying skill depending on seasonality, topography, geography and nature of the rainfall statistic being examined (Awange et al., 2016; Tarnavsky and Bonifacio, 2020). Although some products might outperform each other regionally, there is no widely recognized 'optimal' product across entire continent of Africa, which could be automatically chosen as a validation dataset. This study has therefore focused on two widely available products, cognizant that even as a validation dataset they are likely to have biases. Specifically, we have chosen to focus on the NOAA Rainfall Estimate Version 2 (RFE2) product and the NOAA Africa Rainfall Climatology Version 2 (ARC2). Both have a spatial resolution of 0.1 degrees and a daily temporal resolution.

As described in Xie and Arkin (1996), the Climate Prediction Center's NOAA RFE2 is available from January 2001, with an algorithm incorporating information from gauge data, geostationary infrared, and polar orbiting microwave SSM/I and AMSU-B satellite data (Love et al., 2004; Novella and Thiaw, 2013). RFE2 has been shown to perform well across Africa (Thiemig et al., 2012), as well as in regional specific validations from Egypt ( Abd Elhamid et al., 2020), Burkina Faso (Dembélé & Zwart, 2016), Ethiopia (Gebremicael et al., 2017) and Zimbabwe (Dinku et al., 2018).

NOAA ARC2 is available from 1983 and is derived from a geographically static simple linear relationship between 'cold cloud duration' (the length of time a cloud is colder than a pre-defined threshold of 235K) and historical rain-gauge amounts. This is then merged with rain-gauge data from the World Meteorological Organization's GTS network (Novella and Thiaw, 2013). ARC2 has a well recorded bias in historical trend analysis (Cattani et al., 2020; Maidment et al., 2017) and has been shown to underperform other comparable satellite rainfall products such as CHIRPS or TAMSAT (Satgé et al., 2020; Tarnavsky & Bonifacio, 2020), especially in areas of complex terrain (Ayehu et al., 2018; Dinku et al., 2018). It has been included in this study because it is commonly used by end-users across weather risk management (Osgood & Shirley, 2012) and disease modeling (Kim et al., 2019; Ngwa et al., 2016; Ssentongo et al., 2018). The inclusion of ARC2 also allows this study to be compared directly with other African NWM forecast validations such as Ogutu et al. (2017).

In this study, we take the RFE2 and ARC2 gridded satellite rainfall estimates as the observation data to verify the NWM predictions. The main results of NWM forecasts against the RFE2 rainfall are shown in the main article figures, while the results against the ARC2 are shown in the supplemental figures. Further information about the datasets is shown in Table 1.

**2.2 NWM forecasts**
The ECMWF) and NCEP-GFS are two of the most common products used in global and regional weather and climate forecasting, hydrology, public health, and other applications (Elless & Torn, 2019; Karrouk, 2019; Kerns & Chen, 2014; Meynadier et al., 2010; Nikulin et al., 2012; Sajadi et al., 2020).

The ECMWF's Integrated Forecasting System is used to predict weather and climate using a high resolution (9km) global model and data assimilation system, and has demonstrated considerable skill for synoptic-scale variables in the medium-range in the mid-latitudes (Buizza and Leutbecher, 2015; Park et al., 2008). At the National Centers for Environmental Prediction, the GFS, initialized with the Global Data Assimilation System, creates 16-day predictions four times daily, and output from this model is made available freely around the world. Both data assimilation systems ingest millions of observations from satellite, airborne, ground-based, and ocean-based observing platforms to provide initial conditions for the modeling systems (Cucurull et al., 2007; Kleist et al., 2009), and atmospheric conditions are projected forward in time using the set of equations encoded in the model's dynamical core and physical parameterization schemes. Both the ECMWF and GFS forecasts used in this study are initialized at 12:00 UTC (Table 1). Previous studies have reported that the NCEP-GFS has an advantage over the ocean regions, while the ECMWF products are more reasonable over the Northern Hemisphere continents (Bosilovich et al., 2008; Taraphdar et al., 2016). This study further explores their performances over the African continent.

Table 1. Main characteristics of the satellite and forecast rainfall data used in this study.

| Name | Source | Availability from indicated source | Initialization Time (UTC) | Spatial resolution | Temporal resolution |
|---|---|---|---|---|---|
| RFE2 | ftp://ftp.cpc.ncep.noaa.gov/fews/fewsdata/africa/rfe2/ | 2001-present | N/A | 0.1°x0.1° | Daily Since 12Z |
| ARC2 | ftp://ftp.cpc.ncep.noaa.gov/fews/fewsdata/africa/arc2/ | 1983-present | N/A | 0.1°x0.1° | Daily Since 12Z |
| GFS | https://rda.ucar.edu/datasets/ds084.1/ | 2015-present | -0000 -0600 -1200 -1800 | Native ~13km Interpolated 0.5°x0.5° | 6-hours |
| ECMWF | http://apps.ecmwf.int/datasets/data/tigge/levtype=sfc/type=cf/ | 2006-present | -0000 -1200 | Native ~9km Interpolated 0.5°x0.5° | 6-hours |

## 3. Methods

## 3.1 Verification Scale

At 0.1 degrees, the RFE2 and ARC2 satellite rainfall products have finer spatial resolution than GFS or ECMWF. In order to better compare the rainfall products and forecasts, the satellite observations were rescaled to the model spatial grid of 0.5 degrees. Meanwhile, the temporal resolutions of NWM forecasts are 6-hours, finer than the rainfall estimates of daily resolution, thus we summed the NWM forecast rainfall to a daily time-scale. At the daily aggregation scale, 1-day lead time NWM forecasts (daily forecast sum) were compared to daily resolution satellite observations valid for the same time period (i.e., the days for which the forecasts were predicted match the days the rainfall was observed). At the weekly aggregation scale, NWM forecasts for days 1-7 were summed together, and this 1 week cumulative prediction estimate was compared against the observed 1 week cumulative rainfall estimate valid for the same time period.

## 3.2 Verification Metrics

Several widely used statistical scores were applied to evaluate the NWM forecast performance against the satellite observations, including subtractive Bias, Root Mean Squared Error (*RMSE*), Pearson's Correlation Coefficient (*r*), Accuracy (*A*), Threat Score (*TS*), Probability of Detection (*POD*), False Alarm Ratio (*FAR*), and Brier Score (*BS*). These are summarized further below and in Table 2.

The skill of ECMWF and GFS forecast rainfall in these metrics were compared against the RFE2 and ARC2 satellite rainfall estimates. The Correlation coefficient *r* was applied to measure the strength of the linear relationship between the NWM forecasts and satellite observations. The significance of this result was assessed using a p-value, following a Student's t-distribution test, representing the probability that *r* between the forecast and observations occurred by chance. In this study, *r* is treated as significant when the effective day number (defined as the number of days when there are no missing values in both NWM and satellite data) is larger than 20 and the absolute p-value is less than 0.01; in other words, when a positive correlation between model forecast and satellite observation is of sufficient magnitude to be judged as skillful.

Dichotomous metrics were used to evaluate binary (yes/no) forecasts of rainfall that exceed a threshold amount. It is important to note that there is a complex relationship between NWM forecast error and the values of dichotomous metrics, in part due to the inherent loss of information when dichotomizing a continuous variable such as rainfall (Tartaglione, 2010). Acknowledging this, we include these commonly used metrics to allow intercomparison between other NWM forecast and precipitation validation studies (for example McBride et al 2000; Kidd et al. 2012; Pennelly 2014) and to provide some insight into the ability of the models to capture rain/no-rain. Accordingly, a daily threshold of 2 mm was applied to denote the difference between rain and no rain. This is a commonly used threshold used in rain/no rain impact modelling across sub-Saharan Africa chosen as rainfall amounts less than 2 mm are likely to evaporate rather than be regarded as "useful" rain (Bennett et al., 2011; Mupangwa et al., 2011).

Specific event-based statistics applied include the dichotomous-accuracy, *A*, which is simply the fraction of satellite observed events (either rain or dry) "correctly forecast" in NWM predictions. By correctly forecast, we mean that they correspond with the relevant satellite rainfall estimate rather than against a 'perfect' rainfall observation. The Threat Score, *TS*, represents a somewhat more balanced score of NWM forecast predictions as it removes the cases when no rain was forecast, and no rain was observed (which in some regions represents a majority of cases). The False Alarm Ratio, *FAR*, explains the fraction of NWM forecast events that did not occur in the relevant satellite products. The Probability of Detection, *POD*, compares true positives to all days when rain was observed in the satellite product. Finally, the Brier Skill Score, *BS*, was used to indicate the magnitude of probability forecast errors.

Table 2. Verification metrics list. All metrics are calculated across time (using index *i*, where *N* is the total number of times in the dataset), separately at each spatial location (grid cell). TP: True Positive, hit (event was forecast and occurred in observations). FP: False Positive, false alarm (event was forecast, but did not occur in observation). TN: True Negative, correct negative (event was not forecast and did not occur in observation). FN: False Negative, missed event (event was not forecast but did occur in observation). $RF_i$: rainfall forecast from NWM (unit: mm day$^{-1}$) on day *i*. $RS_i$: rainfall observation from satellite (unit: mm day$^{-1}$) on day *i*. $F_i$: the probability that was forecast on day *i*. $O_i$: the actual outcome that was observed on day *i*.

| Statistical Index | Equation | Range | Perfect Value |
|---|---|---|---|
| Bias (subtractive) | $BiasA = mean(RF_i - RS_i)$ | $-\infty$ to $\infty$ | 0 |
| Root Mean Squared Error (RMSE) | $RMSE = \sqrt{\frac{1}{N}\sum(RF_i - RS_i)^2}$ | $0$ to $\infty$ | 0 |

| Pearson's Correlation Coefficient (r) | $r = \dfrac{\sum(RF_i - \overline{RF})(RS_i - \overline{RS})}{\sqrt{\sum(RF_i - \overline{RF})^2 \sum(RS_i - \overline{RS})^2}}$ | -1 to 1 | 1 |
|---|---|---|---|
| Accuracy (A); Proportion Correct | $A = \dfrac{TP + TN}{TP + FP + FN + TN}$ | 0 to 1 | 1 |
| Threat Score (TS); Critical Success Index | $TS = \dfrac{TP}{TP + FP + FN}$ | 0 to 1 | 1 |
| False Alarm Rate (FAR) | $FAR = \dfrac{FP}{TP + FP}$ | 0 to 1 | 0 |
| Probability Of Detection (POD) | $POD = \dfrac{TP}{TP + FN}$ | 0 to 1 | 1 |
| Brier Score (BS) | $BS = \dfrac{1}{N}\sum(F_i - O_i)^2$ | 0 to 1 | 0 |

### 3.3. Regions of Interest

In order to more clearly understand spatial differences in NWM forecast performance over Africa, we calculated validation metrics separately over Northern Africa (NA, land: 19°W-40°E, 20°N-36°N), Western Africa (WA, land: 19°W-10°E, 0°N-20°N), Eastern Africa (EA, land: 30°E-51°E, 10°S-20°N), Central Africa (CA, land: 10°E-40°E, 10°S-20°N) and Southern Africa (SA, land: 10°E-40°E, 36°S-10°S), as indicated in Figure 1. All evaluations are performed over the African landmass, while the ocean is masked.

## 4. Results

### 4.1 Spatial and Seasonal Pattern

African annual rainfall shows considerable spatial variation, with the largest climatological values of precipitation concentrated around the equatorial region and dominated by the Intertropical Convergence Zone, ITCZ (Figure 1). Figure 2 depicts the spatial pattern of averaged daily rainfall, broken down by season and product. The NWM rainfall forecasts from both ECMWF and GFS show a a similar overall spatial pattern over Africa for each season, which are similar to those patterns observed by satellite. For example, the general seasonality of the rainfall belt, controlled by the movement of the ITCZ, is captured in its north-south movement by all four products. The differences between west coastal and east coastal Africa also are clearly represented in the NWM forecasts, especially during December-January-February (DJF) and June-July-August (JJA) (Figure 2, columns 1 and 3). It is interesting to note that terrain-enhanced rainfall is more pronounced in GFS compared to ECMWF. For example, Ethiopia's Bale Mountain range (40 E 8 N) in MAM and JJA, and the Marra Mountain range in western Sudan (34 E 13 N) in JJA are clearly visible in the model output.

**4.2 Verification of NWM forecasts at daily scale**

Overall, both NWM and the satellite observations show a similar overall magnitude of rainfall amounts. For example, the spatial mean of total rainfall in three years from 2016-2018 across the whole Africa continent, is very similar across both satellites (ARC2: 1718 mm, RFE2: 1772 mm) and models (ECMWF: 1786 mm, GFS: 1714 mm). Despite this, there are clear spatial differences between the NWP forecasts and satellite observations at daily time-scales. For example, both the ECMWF and GFS forecasts tend to underestimate rainfall totals in tropical Africa while overestimating rainfall totals elsewhere (Figure 3 and Figure S1). The largest (subtractive) bias occurs over and around the ITCZ rainfall region; both ECMWF and GFS tend to have larger biases over West Africa than that in other regions, especially during JJA. The Central and East African highlands are another important region of precipitation bias, shown especially in the GFS forecast.

As shown in Figure 3, ECMWF and GFS differ in their spatial representation of African rainfall, especially over the coastal regions of West Africa. One of the most significant differences is in the placement of the ITCZ during JJA; there is a strong dry bias in the GFS at around 10 N, while the bias is somewhat weaker and placed further north in the ECMWF, with a wet bias appearing around 5 N. A spatial pattern such as this is often observed when two datasets disagree on either the speed or spatial extent of the ICTZ movement; for example, in this case it appears that both models underestimate the Northern extent of timing of the West African monsoon. This corresponds with previous research on both the ECMWF family of products (ECMWF System 3: Tompkins and Feudale, 2010) and the GFS family of products (Xue et. al 2010), which found NWM have a tendency to displace the rainfall in the West African Monsoon too far south. These discrepancies might be partially attributable to errors in convective parameterization schemes. This conclusion is also supported by the fact that convection systems contribute more than 50% of the annual rainfall over the west coast equatorial region (Maranan et al., 2018), but are difficult to predict well without a convection-allowing (grid spacing 3km or smaller) model (e.g. Clark et al., 2016).

Our results also indicate that complex topography and terrain makes an important contribution to the NWM forecast bias. Terrain is conducive to generating lift, and therefore triggering convection (as evidenced by enhanced rainfall over terrain in East Africa and Madagascar). The GFS is also more sensitive to topographically enhanced precipitation, showing a wet bias relative to satellite estimates. Additionally, biased water vapor measurements and sparsity of available observational data are likely contributing factors to the forecast biases, as poor initial conditions lead to deficiencies in model depictions of rainfall (Jiang et al 2013). It is likely that these measurement quality factors could also contribute to the different performances among different numerical weather models.

When considering RMSE at the daily scale, the timing of precipitation becomes more important rather than just the seasonal totals used to calculate the bias. In Figure 4 and Figure S2, higher values of RMSE were centered on the ITCZ rainfall regions during each season. The GFS model has larger RMSE values than the ECMWF, especially over the ITCZ rainfall regions. While the models have identified the general region and seasonality of the ITCZ, it is especially challenging to predict the exact day, location, and amount of rainfall from local convection. Large scale NWM forecast rainfall spatial patterns are controlled by local circulation patterns, such as Land-Based Convergence Zones, which are closely connected with the tropical Atlantic ITCZ in the ECMWF model but more dominated by Indian Ocean ITCZ in the GFS model (Zhang et al., 2013). The degree to which the models correctly depict the amplitude and phase of African Easterly Waves would also be reflected in the RMSE.

As described in the methods, the correlation coefficient ($r$) was used to determine the degree of linear association between NWM forecasts and satellite observations across the temporal dimension. Figure 5 and S3 show spatial distributions of correlation coefficient at locations with positive skill (p-value less than 0.01 and effective day number of 20). These criteria were set to avoid spurious linear correlation values caused by missing information or very high percentage of dry days, especially over North Africa where the NWM forecasts have a very large correlation coefficient but do not reach statistical significance at most locations. The greatest model skill at the daily time scale is found from 0 to 10 N in DJF, and 10-20 S and over East Africa in MAM and SON. In many regions, the NWM forecasts cannot explain the observed rainfall well at the daily scale. This is especially apparent during JJA, as less than 10% of the locations in either forecast have substantial linear correlation with the RFE2 rainfall (Figure 5c and 5g). Furthermore, at each location, less than 50% variation of the observed rainfall could be explained by the NWM forecasts. During March-April-May (MAM), the correlation coefficient in the Southern Africa region is larger than that of other seasons; while in September-October-November (SON), the correlation in the West Africa region is stronger than other seasons. In addition, we observe that the ECMWF forecast shows a stronger correlation with the satellite estimates than the GFS during all four seasons (Figure 5 and S3). The correlation coefficients show an interesting spatial pattern, that does not appear to be fully linked with the progression of the ITCZ. A band of higher correlation coefficients are seen across West Africa during the dry season, but an equivalent band is not seen in the Southern Hemisphere. Very low correlation coefficients are also seen in Western Equatorial Africa, a region dominated by limited observing networks and a complex synoptic situation (Nicholson 2013), suggesting that NWMs are less able to capture the dynamics in that region.

We now turn to dichotomous (hit/miss) statistics to assess the ability of the NWM forecasts to capture the occurrence of rainfall events. The results for ECMWF are depicted in Figure 6. with the results for GFS shown in Figure 7. Overall, many of the dichotomous metrics were dominated by the large-scale climatology of the study area, in part due to the nature of the statistics themselves. For example, for both models, high dichotomous-accuracies, *A*, were recorded in climatologically dry regions, or perpetually dry areas such as the Sahara Desert. In comparison, the general drop in Accuracy in rainy-season areas is due to the fact that during the rainy-season 'a rain day' is less dominant a climatological category than a dry day is in the dry season. For example, there are often days during the rainfall season when it doesn't rain, but typically very few rain-days in the dry season. As the accuracy is therefore heavily influenced by the most common climatological category (rain/no-rain), we include it here as the lowest bar a dichotomous forecast should pass and suggest that the result provides additional confirmation that in general the model is reproducing the large scale movement of the ICTZ.

At the next level of complexity, we study the False Alarm Ratio (FAR), depicted in Figure 6 and Figure7 m-p, and the Probability of Detection (POD), depicted in Figure 6 and Figure 7 i-k.For these statistics, a value of 1 means a high Probability of Detection (good) or a high False Alarm Ratio (bad). In general, these also capture the movement of the ICTZ, although there is evidence that ECMWF is capturing orographic rainfall increases in areas such as the western Ethiopian highlands (JJA and SON). This result has been independently confirmed across several other studies and forecast timescales (for example Ehsan et al (2021) at seasonal timescales). It is also worth noting that this is a complex region where satellite rainfall datasets can also struggle to accurately capture rainfall occurrences (Dinku et al. 2007). Subtle differences are also seen between ECMWF and GFS, for example in South Africa during JJA, GFS captures a different pattern of rainfall events than ECMWF, with ECMWF forecasting broadly more rainfall in the region, leading to more captured rainfall events at the expense of higher false alarms. In general, higher False Alarm Ratios are seen at the edges of the ICTZ rainfall band, adding further evidence that the models might be mischaracterizing the exact location of the ITCZ in a given year. This is likely to be highlighted given the relatively short time-period used in the analysis (2016-2018). The short time period and low threshold for rain/no-rain also means that some data artifacts are apparent; for example, the small area of higher POD in Nigeria in the DJF 'dry season' is likely due to a single rainfall event.

Finally the Threat Score, TS, is shown in Figure 6/7 e-h and the Brier Score, $BS$, shown in Figure 6/7 q-t. The Threat score measures the proportion of correctly predicted forecast events, e.g. the accuracy when correct negatives have been removed, whilst the Brier score, or frequency bias, assesses the magnitude of the probability forecast errors. Both confirm similar spatial patterns to those discussed above, although it is interesting to note the higher skill in East Africa during the early rains (MAM). In general at a daily scale, our results are consistent with several previous regional studies including Kniffka et al., 2020; Camberlin et al., 2019; Vogel et al. 2018; Parker and Diop-Kane, 2017; and Ma et al 2019, which show limited daily model skill. We therefore suggest that although the daily forecasts are able to capture large scale features, if one is interested in precision for the rainfall occurrence at a specific 0.5 degree pixel on a specific day, both ECMWF and GFS still have room for improvement.

It is worth highlighting that through the lens of public health or other applications, even when models struggle to forecast daily pixel-scale rainfall, higher skill is often observed for other convective statistics that could prove useful predictors of public health. For example, Vigaud and Giannini (2019) recorded higher categorical skill scores when assessing ability of ECMWF forecasts to correctly classify a location into one of 7 convective regimes. It should also be noted that the daily dichotomous skill scores in Figure 6 and Figure 7 for rainfall might be depressed due to diurnal reporting mismatch. Both NWM and satellite rainfall observations are poor at predicting the exact hour rainfall occurred (Greatrex et al. 2012). Therefore, given that tropical convective rainfall often falls overnight, it is common to observe a 1-day mismatch in daily rainfall records where one rainfall event is split over two days in one product, but recorded in a single day in another. Equally, cloud features such as cirrus anvils can also lead to reduced satellite rainfall skill at a daily scale compared to temporally aggregated data, and the binary nature of dichotomous hit/miss statistics mean that these mismatches are heavily penalized. In the next section, we therefore consider weekly scale verification to remove these biases.

**4.3 Weekly scale verification and influence of forecast lead time**
As the skill of an NWM forecast is highly dependent on verification time-scale, we assessed NMW forecasts of aggregated rainfall over 1 week compared to daily rainfall with lead time of one day (Figures 3-9). Precipitation biases at a weekly scale were reduced from the 1-day forecasts (Figure 3), reflecting the impact of model initialization and forecast spin-up in developing convective systems. We found that at a weekly time-scale, both the ECMWF and GFS forecasts show better performances than those at a daily time-scale. For example, weekly forecasts show lower RMSE (Figure 4) and higher correlation coefficients (Figure 5). The improvements in RMSE were especially prevalent over areas dominated by the ITCZ; the more frequent the rainfall events, the more likely that aggregating to a one-week time-scale improves performance. For correlation coefficient, the largest improvements were at the edges of areas of rainfall; north of the equator in DJF, in southern Africa and East Africa in MAM and SON; skill remained low in JJA. Differences in dichotomous statistics for the ECMWF (Figure 8) and the GFS (Figure 9) share a similar spatial pattern. Improvements in accuracy followed the seasonal movement of the ITCZ; regions in their rainy season often exceeded the threshold precipitation value for weekly precipitation in both model and observations, making this an easier forecast than at the daily scale where this may not happen every day at a particular location. Likewise, threat score and probability of detection also improved over those regions (particularly Central and Southern Africa). For False Alarm Rate, improvements were more modest and did not reveal a clear spatial pattern. For brier score, improvements (reductions in score) also followed the ITCZ and regions of improved accuracy.

NWM performance also depends on forecast lead time (e.g. Greybush et al., 2017). While the models capture seasonal changes in rainfall, it is difficult to predict weather and climate anomalies at long lead times of weeks or months due to the chaotic nature of the atmosphere (Luo and Wood, 2006). At shorter sub-daily time ranges, NWM forecasts would need to capture the location of convection elements, or individual thunderstorm clusters. Therefore, the differences between NWM and satellite datasets would increase when they are evaluated on a finer time scale or at a longer lead time. For rainfall, it appears the benefit of aggregating rainfall over a 7 day period outweighs forecast degradation at lead times extending from 1 day to 7 days.

The results suggest that as suggested in the previous section, improvements in model skill might be achieved through less emphasis on the exact time of an individual rainfall event. This has important implications for end-user applications, especially in agriculture and health, as the majority of such uses are not reliant on knowing the precise rainfall at a specific time on a specific day but rather the cumulative impacts of an excess or deficit of rainfall. We suggest that 1-week aggregated NMW forecasts show potential in supporting public health and other end-user needs.

### 4.4 Differences of rainfall intensity and frequency between forecast and satellite

Rainfall intensity and frequency are key factors of the temporal and spatial variations shown in previous figures. Therefore, Figure 10 compares the relative distribution of events (cumulative number of days over all locations) falling in each of six daily precipitation intensity categories between RFE2 and ECMWF, separated by season. The lowest intensity events (less than 2 mm day$^{-1}$, which count as no rain for the dichotomous statistics) comprise the greatest number of days, with more such days in JJA and SON for the EMWF compared to RFE2 in MAM, JJA, and SON. This pattern flips for events of 2-10 mm day$^{-1}$, with RFE2 having more such events in all seasons compared to ECMWF. As the rainfall threshold increases, events become less common. For rainfall of more than 20 mm day$^{-1}$, there are clearly many more model-predicted events than observed events. This figure shows that there are more low intensity (less than 2 mm day$^{-1}$) and high intensity (>20 mm day$^{-1}$) rainfall events forecasted by the NWM than observed, which has importance if the forecasting application depends upon the amount of rainfall.

### 5. Discussion and Conclusions

This study examined the skill of NWM forecasts relative to satellite rainfall observations over Africa during three years: 2016-2018. It was shown that two NWM forecasts, ECMWF and GFS, show large spatial and seasonal variabilities in skill over the African continent. In terms of seasonal climatology, the NWMs generally reproduced observed amounts and locations, with some discrepancies in location of the ITCZ and amount of orographic precipitation. On the shorter time scale, ECMWF forecasts demonstrated better performance than GFS forecasts in terms of lower RMSE and FAR values and higher accuracy and correlation coefficient. As expected, we found that a weekly aggregation time compared to a daily aggregation time exhibited higher correlation, threat score, and POD, and lower RMSE values. These results show that NWM forecasts do have some positive skill relative to satellite observations, but that there is substantial room for improvements in precipitation forecasting. This could be obtained via better observations, data assimilation techniques, and model resolution and parameterizations (such as convection parameterizations). Model skill over Africa is lower than in other parts of the world with more dense observing systems, such as North America (Yang et al., 2017). In addition, the use of regional high resolution, convection-allowing models (with grid spacings of 3km or less) may better resolve local mesoscale weather phenomena, such as lake breeze convection, orographic convection, the diurnal cycle of convection, and other mesoscale convective systems, leading to improvement in predictive skill of precipitation.

Potential limitations arise from the focus constraint on the years 2016-2018. This three-year period may exclude certain extreme weather and climate events that could have led to different quantitative metrics of prediction skill. They also mean that especially in the dry season areas, some validation features might be the result of very few rainfall events. The advantage of using this time period is its relative recency, allowing us to only validate modern NWM forecasts, observations, and assimilation systems, as well as alignment with the needs of ongoing public health projects. Considering these advantages and limitations, we expect the broad conclusions of this paper to be applicable to other years not included in the study.

The results of this study have important implications for public health P3H efforts. We have increasing evidence that certain soil bacteria, such as the Burkholderia that causes Melioidosis in Southeast Asia and Northern Australia (Wiersinga et al., 2018), and perhaps the Paenibacillus that causes infant brain infections and postinfectious hydrocephalus in East Africa (Paulson et al 2020; Schiff et al 2012), may be environmentally influenced by rainfall. If soil moisture is important in regulating the infectivity of soil and drinking water, then fusing public health microbial surveillance with satellite environmental measurements and NWM forecasts form a natural combination. Such fusion offers the prospect of real-time prediction of the likely organisms infecting patients who present with signs and symptoms of serious bacterial infections. Causative organism prediction is critical in the proper selection of antibiotics. In low- and middle-income countries, where availability of microbiological laboratory testing is limited, predictive methods may offer immediacy in guiding antimicrobial treatment alternatives. Furthermore, P3H efforts fusing historical microbial surveillance patterns with environmental factors can be utilized in optimized prevention efforts focusing attention on places and timing where the anticipated infection risk is greatest. Finally, bacteria in soil, or arthropod vectors breeding in water, require relevant time scales to increase their density. Such time scales are typically longer than individual days, and likely extend to one or more weeks. Therefore our demonstration in this present study of increased forecasting skill utilizing 1-week cumulative rainfall is of particular interest in further exploration of P3H.

**6. Acknowledgements:** We dedicate this work to the memory of Professor Fuqing Zhang, our colleague who was vital to the concept and initiation of this project. The authors declare no conflicts of interest. This work was supported by an NIH Director's Transformative Award 1R01AI145057.

**7. Data Availability**
GFS and ECMWF NWM data are available for download at the UCAR (https://rda.ucar.edu/datasets/ds084.1/) and TIGGE (http://apps.ecmwf.int/datasets/data/tigge/levtype=sfc/type=cf/) websites, respectively. RFE2 (ftp://ftp.cpc.ncep.noaa.gov/fews/fewsdata/africa/rfe2/) and ARC2 (ftp://ftp.cpc.ncep.noaa.gov/fews/fewsdata/africa/arc2/) datasets are available from the NOAA NCEP Climate Prediction Center website.

**8. Code Availability**
The code required to replicate the findings in this paper are archived at: https://github.com/Schiff-Lab/African-Rainfall

**9. Supporting Information**
Figures S1 (ECMWF and GFS bias comparison with ARC2), S2 (ECMWF and GFS RMSE comparison with ARC2), and S3 (ECMWF and GFS correlation coefficient comparison with ARC2) are included in the supporting information, and are referenced in the text.

Yang, X., R. Siddique, S. Sharma, S. J. Greybush, and A. Mejia, 2017: Postprocessing of GEFS precipitation ensemble reforecasts over the U.S. middle-Atlantic region. Mon. Wea. Rev., 145, 1641-1658, doi:10.1175/MWR-D-16-0251.1.

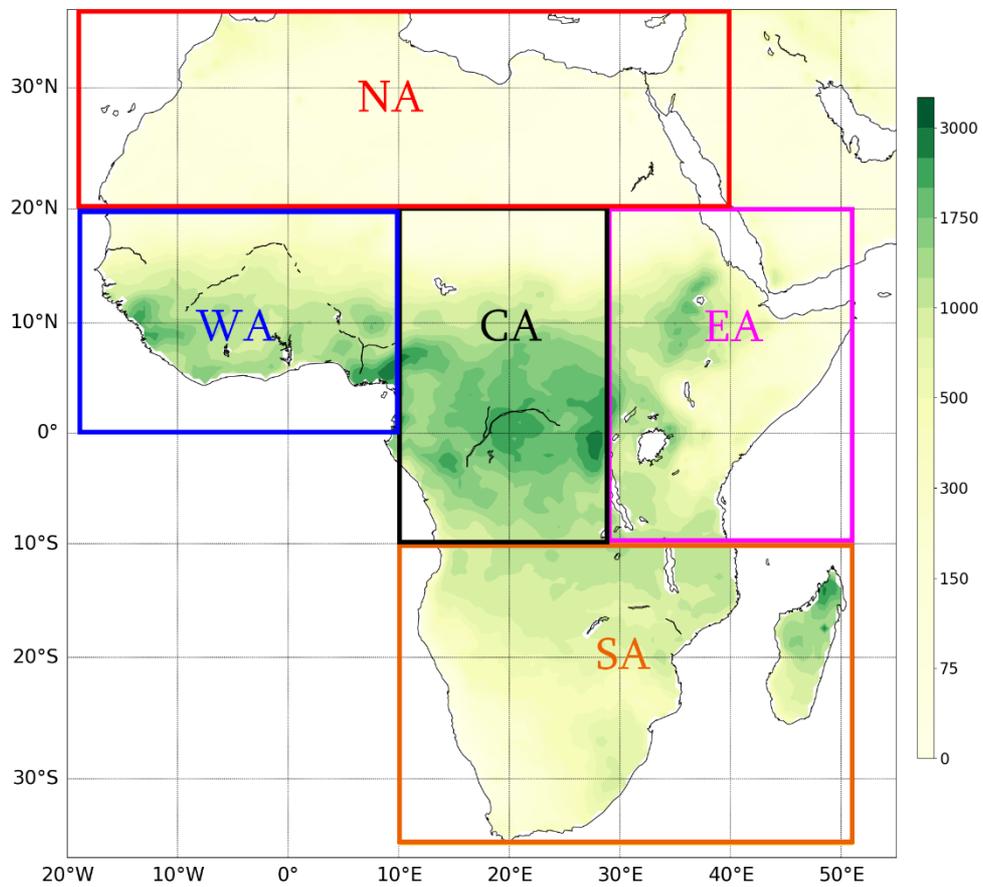

Figure 1. Regions of Africa used in this study: NA: Northern Africa; WA: Western Africa; CA: Central Africa; EA: Eastern Africa; SA: Southern Africa. Background image is spatial pattern of total RFE2 rainfall (units: mm), summed over 2016-2018.

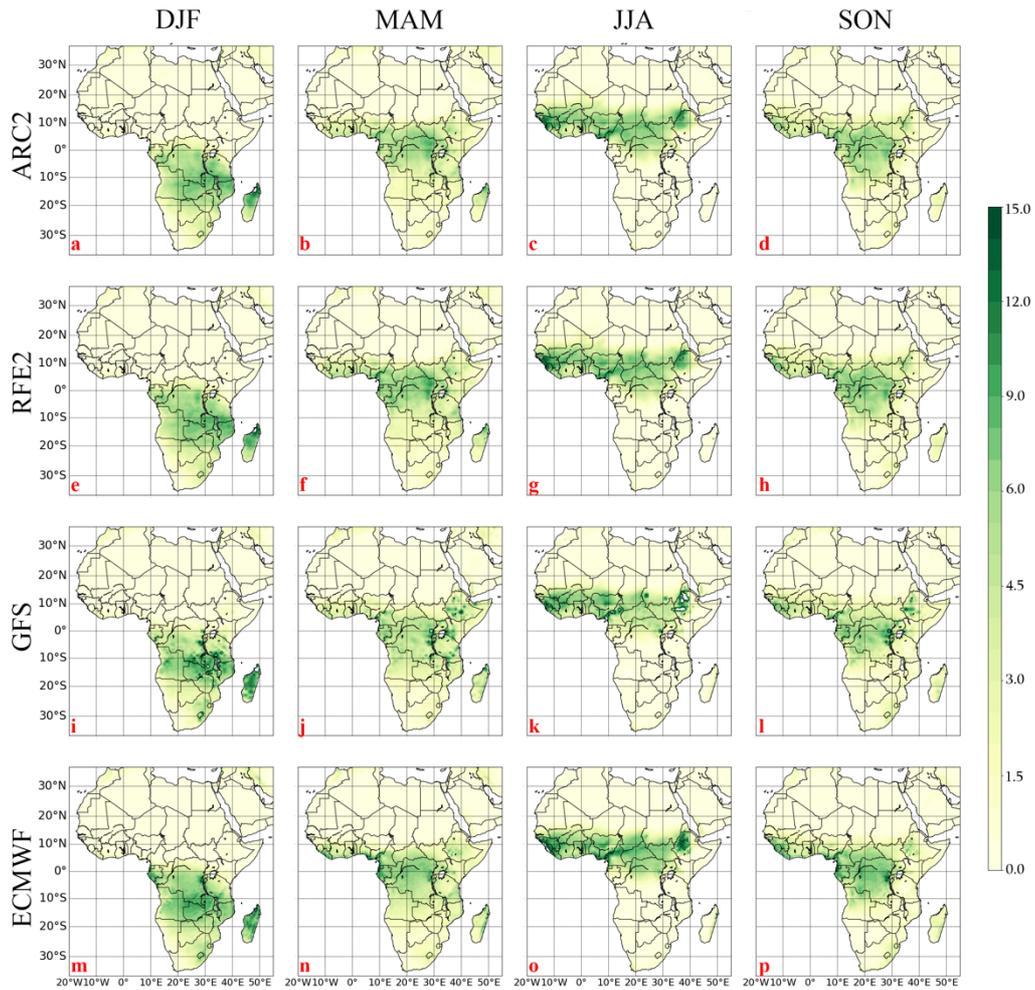

Figure 2. Spatial pattern of averaged daily rainfall, averaged during 2016-2018 (units: mm day$^{-1}$) of ARC2 (a, b, c, d), RFE2 (e, f, g, h), GFS (i, j, k, l), and ECMWF (m, n, o, p) for each season (December-January-February, DJF: a, e, i, m; March-April-May, MAM: b, f, j, n; June-July-August, JJA: c, g, k, o; September-October-November, SON: d, h, l, p). Note: the ECMWF and GFS forecasts are at lead time of one day.

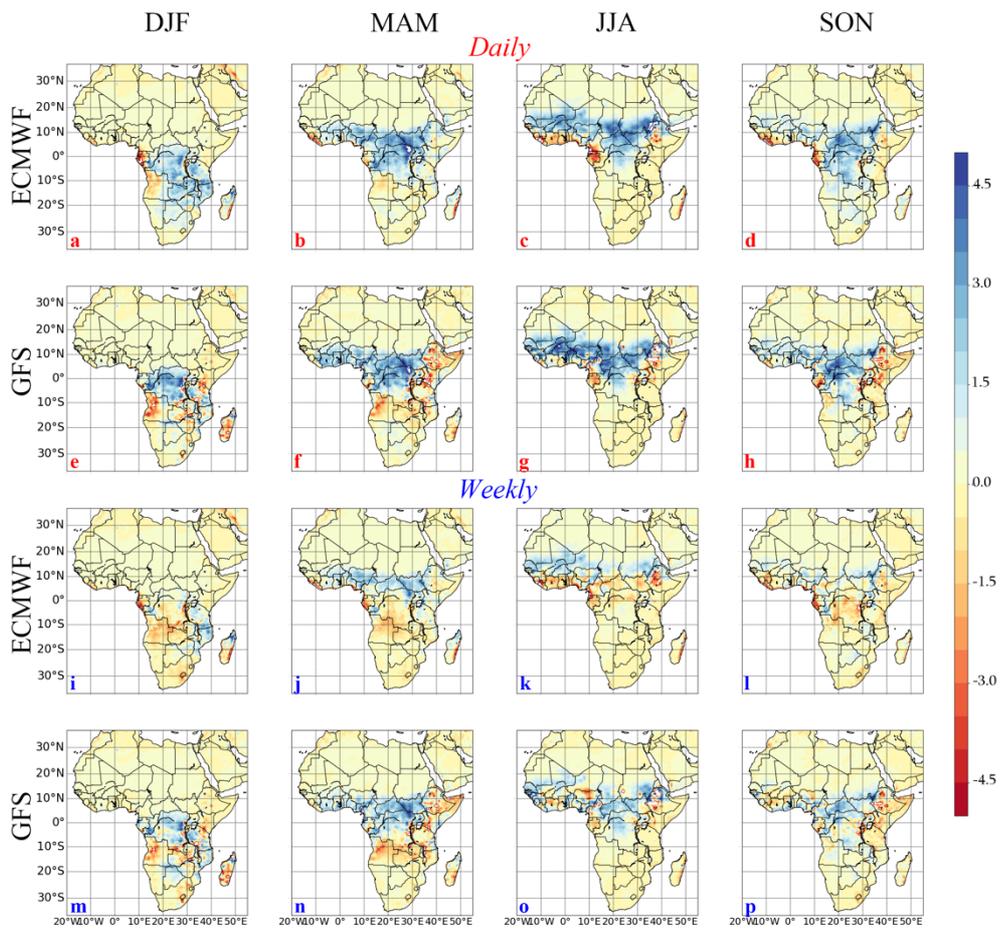

Figure 3. Spatial distribution of Bias (mean observation minus model; units: mm day$^{-1}$) between NWM forecasts (ECMWF and GFS) and RFE2 observations, at daily and weekly scales, in each season. Note: the NWP forecasts for daily rainfall prediction are at one day lead time.

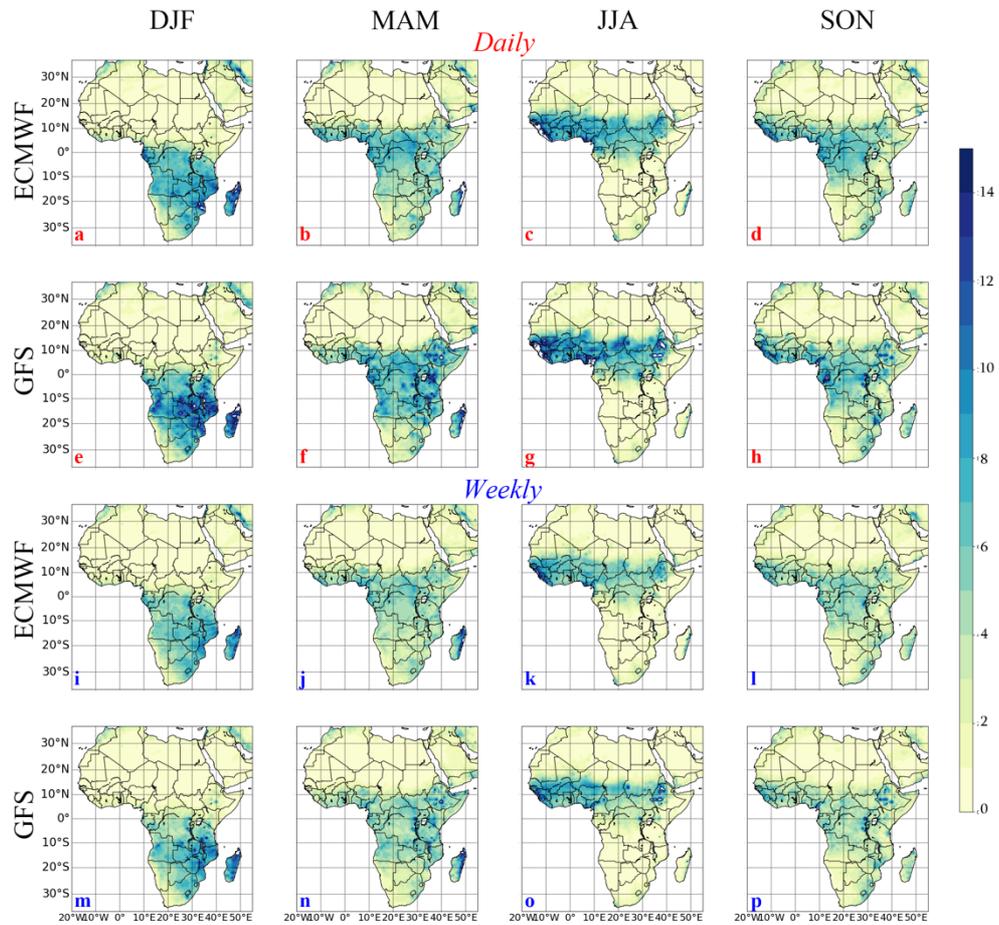

Figure 4. Spatial distribution of RMSE (units: mm day$^{-1}$) between NWM forecasts (ECMWF and GFS) and RFE2 observations, at daily and weekly scales, in each season. Note: the NWP forecasts for daily rainfall prediction are at one day lead time.

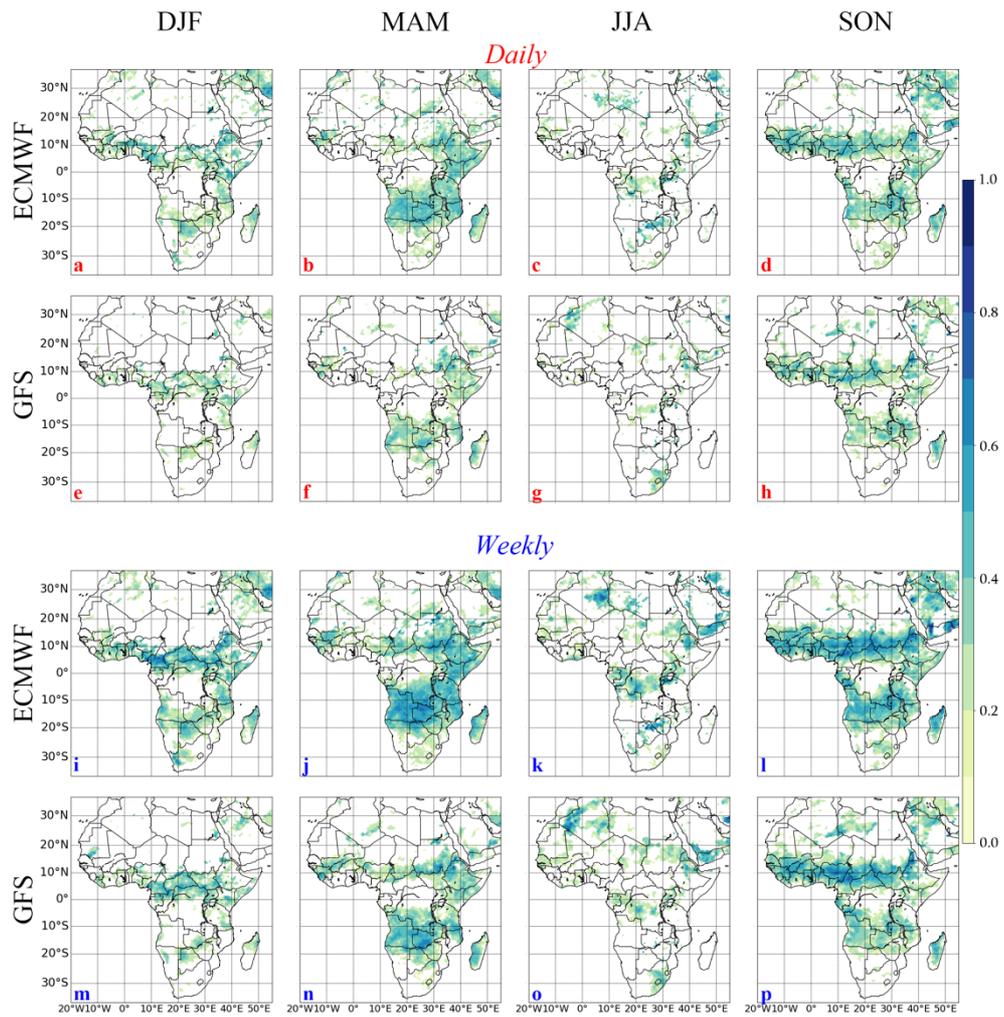

Figure 5. Spatial distributions of Correlation Coefficient (r) between NWM forecasts (ECMWF and GFS) and RFE2 observations, at daily and weekly scales, in each season. Note: the NWP forecasts are at one day lead time, and locations with effective days > 20 and P value < 0.01 (indicating positive forecast skill) are shown in color, hence the colorbar with positive values only.

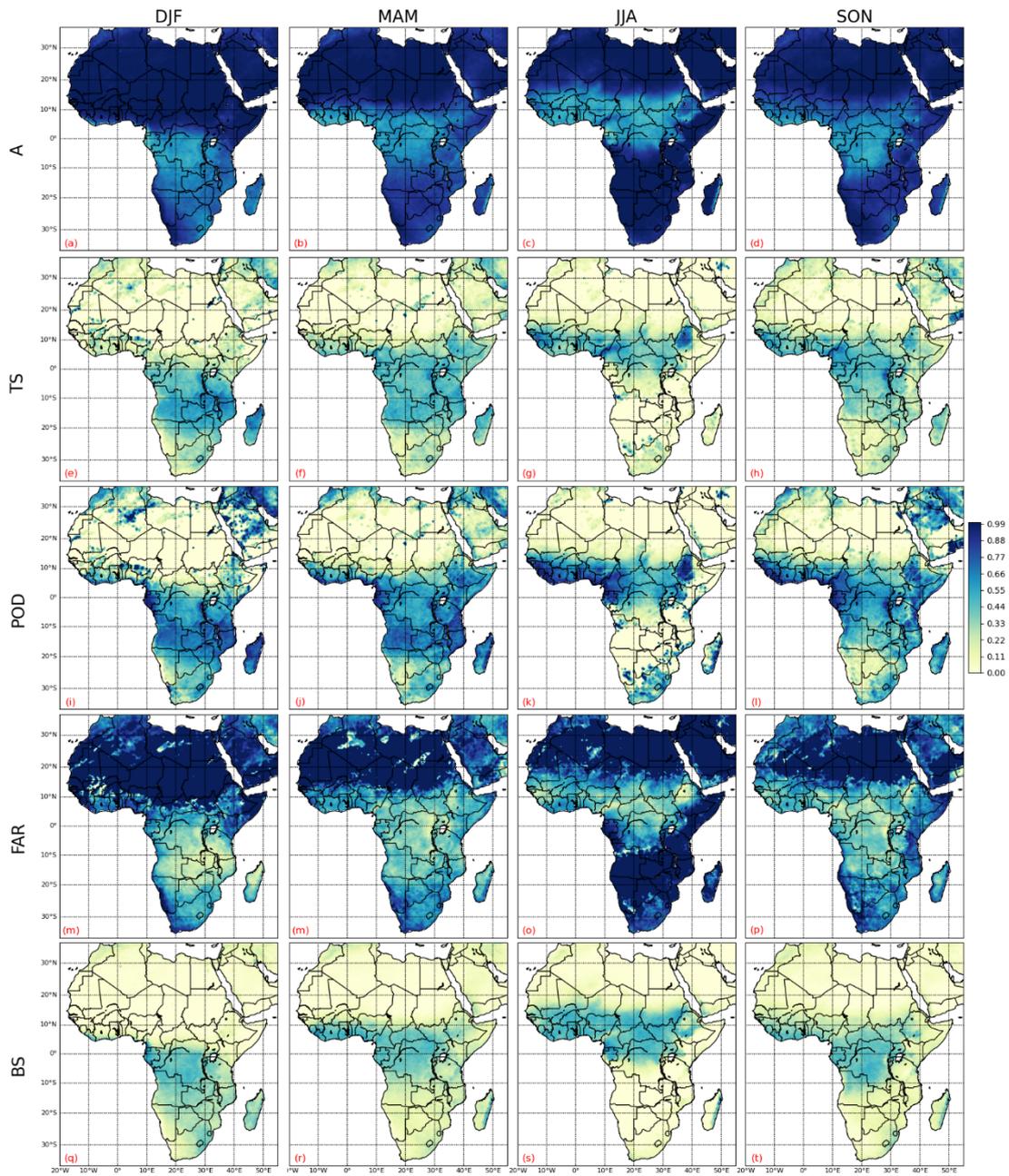

Figure 6. At daily scale, spatial distributions of Accuracy (a, b, c, d), Threat Score (e, f, g, h), Probability Of Detection (i, j, k, l), False Alarm Rate (m, n, o, p) and Brier Score (q, r, s, t) of ECMWF forecasts at one day lead time against the RFE2, in each season, during 2016-2018. Note: threshold is 2 mm day$^{-1}$

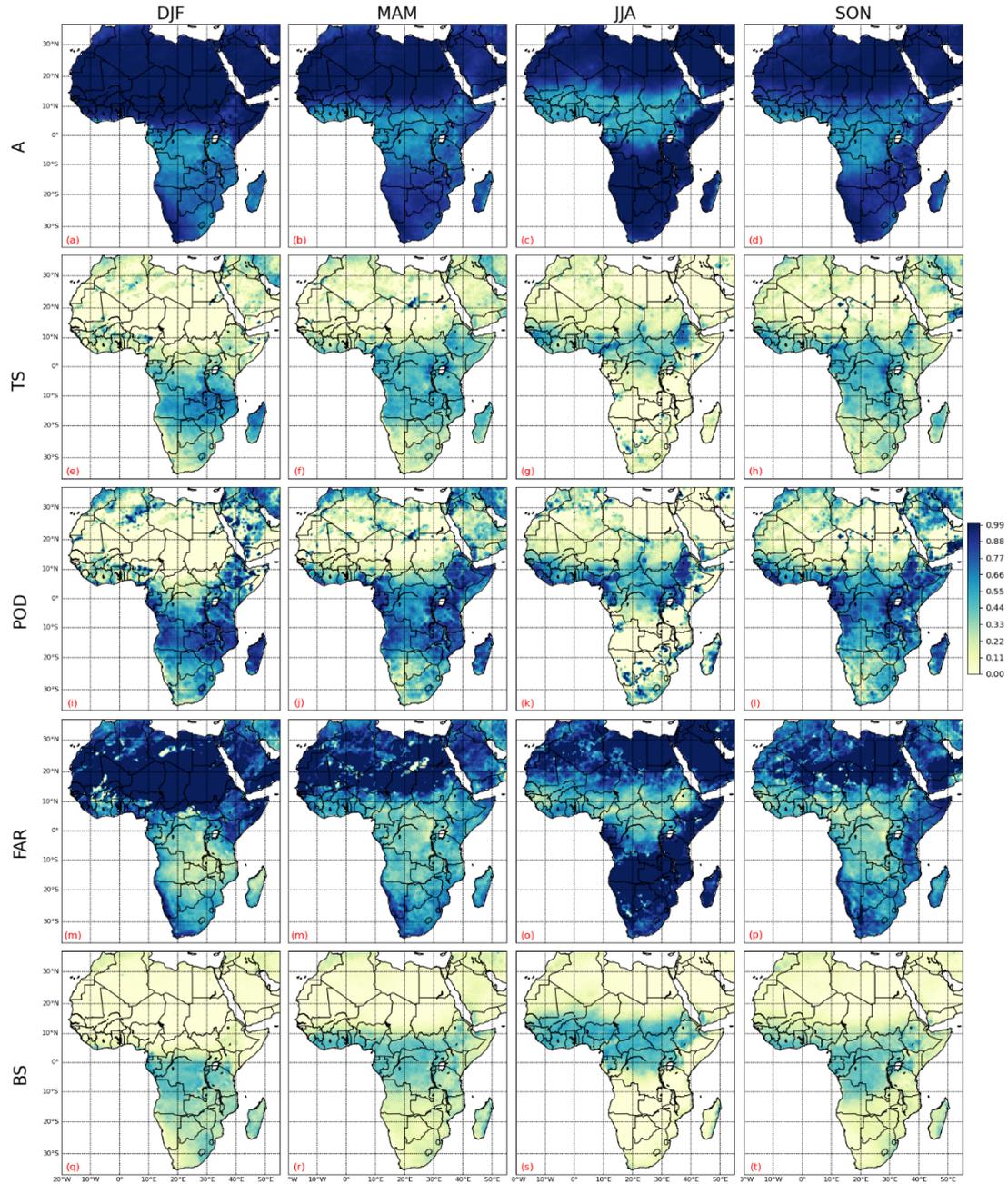

Figure 7. At daily scale, spatial distributions of Accuracy (a, b, c, d), Threat Score (e, f, g, h), Probability Of Detection (i, j, k, l), False Alarm Rate (m, n, o, p) and Brier Score (q, r, s, t) of GFS forecasts at one day lead time against the RFE2, in each season, during 2016-2018. Note: threshold is 2 mm day$^{-1}$

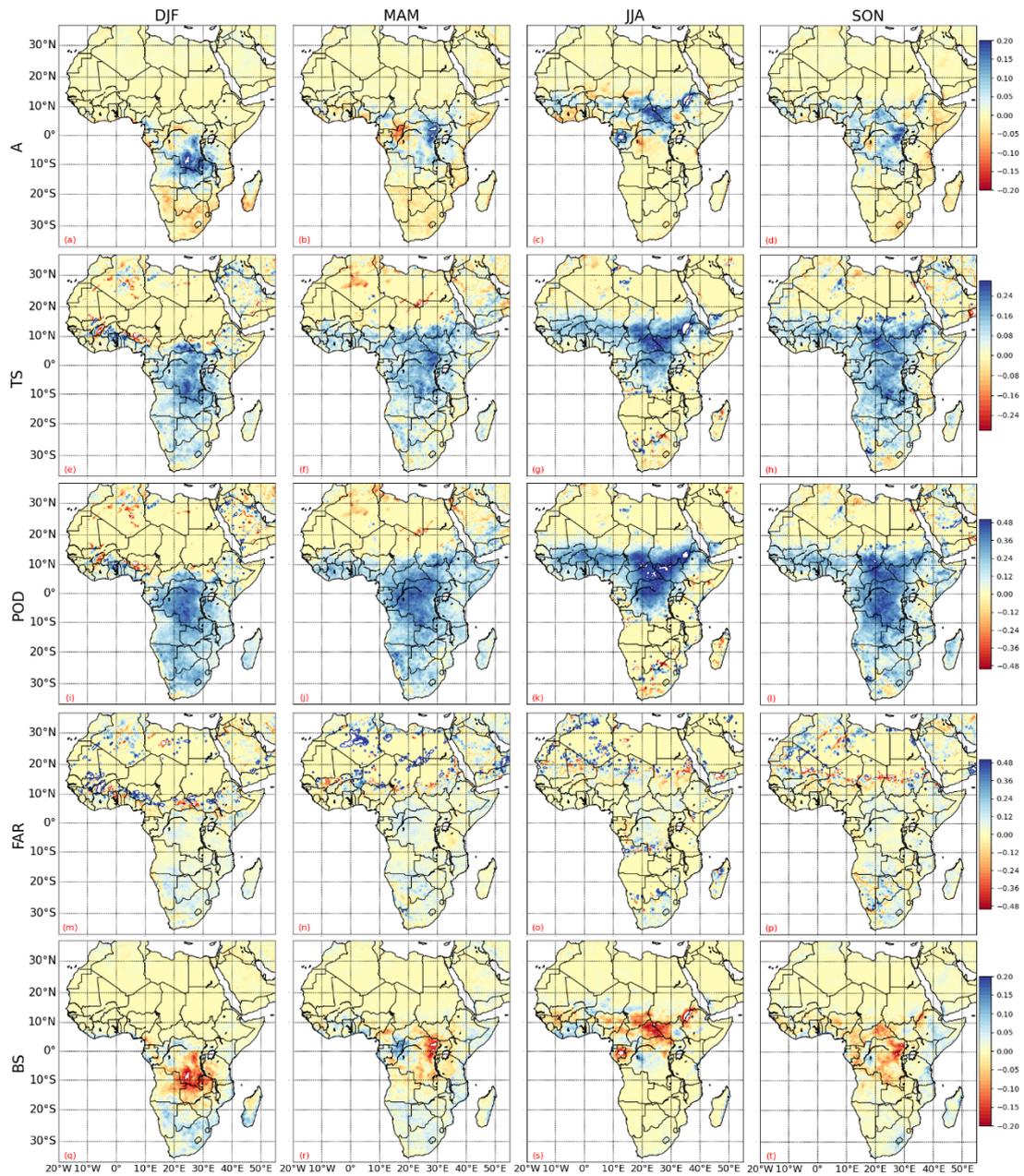

Figure 8. Weekly NWM forecast metrics minus daily forecast metrics. Note: the ECMWF NWM forecasts for daily rainfall prediction are at one day lead time. A positive value (blue) indicates an improved performance for weekly versus daily for A, TS, and POD; a negative value (red) indicates an improved performance for weekly versus daily for FAR and BS.

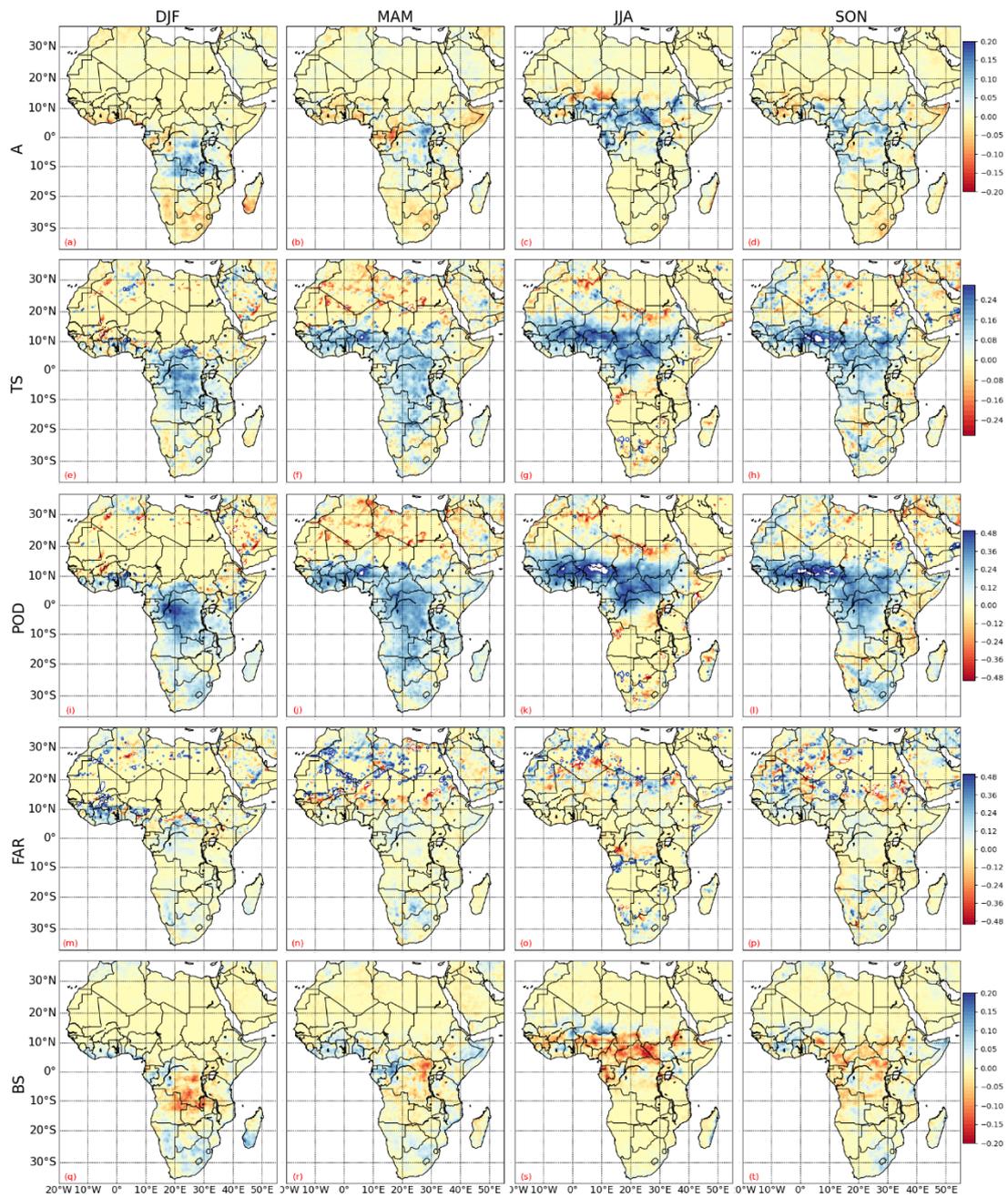

Figure 9. Weekly NWM forecast metrics minus daily forecast metrics. Note: the GFS NWM forecasts for daily rainfall prediction are at one day lead time. A positive value (blue) indicates an improved performance for weekly versus daily for A, TS, and POD; a negative value (red) indicates an improved performance for weekly versus daily    for FAR and BS.

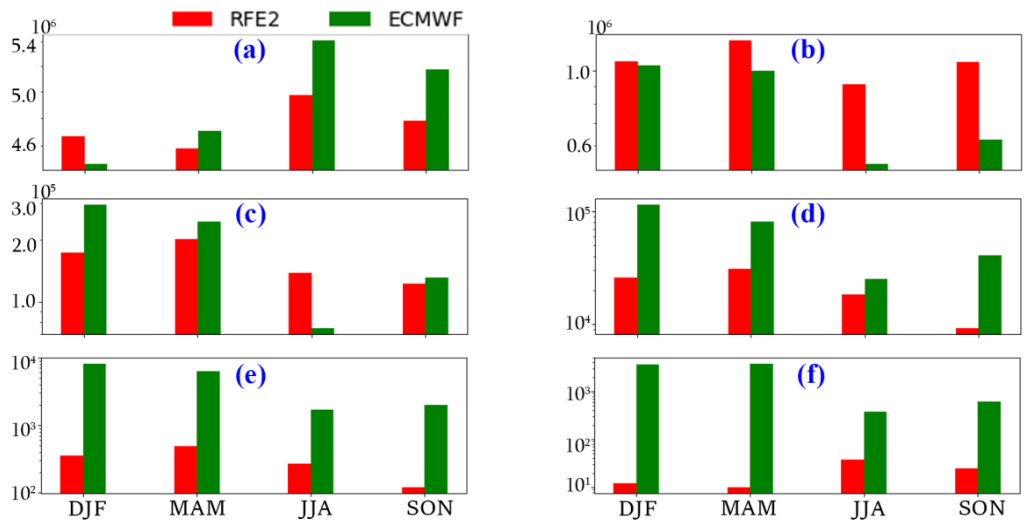

Figure 10. Cumulative number of days over all locations per season for RFE2 and ECMWF rainfall in different intensity ranges: (a) 0-2 mm day$^{-1}$, (b) 2-10 mm day$^{-1}$, (c) 10-20 mm day$^{-1}$, (d) 20-50 mm day$^{-1}$, (e) 50-75 mm day$^{-1}$, and (f) > 75 mm day$^{-1}$.

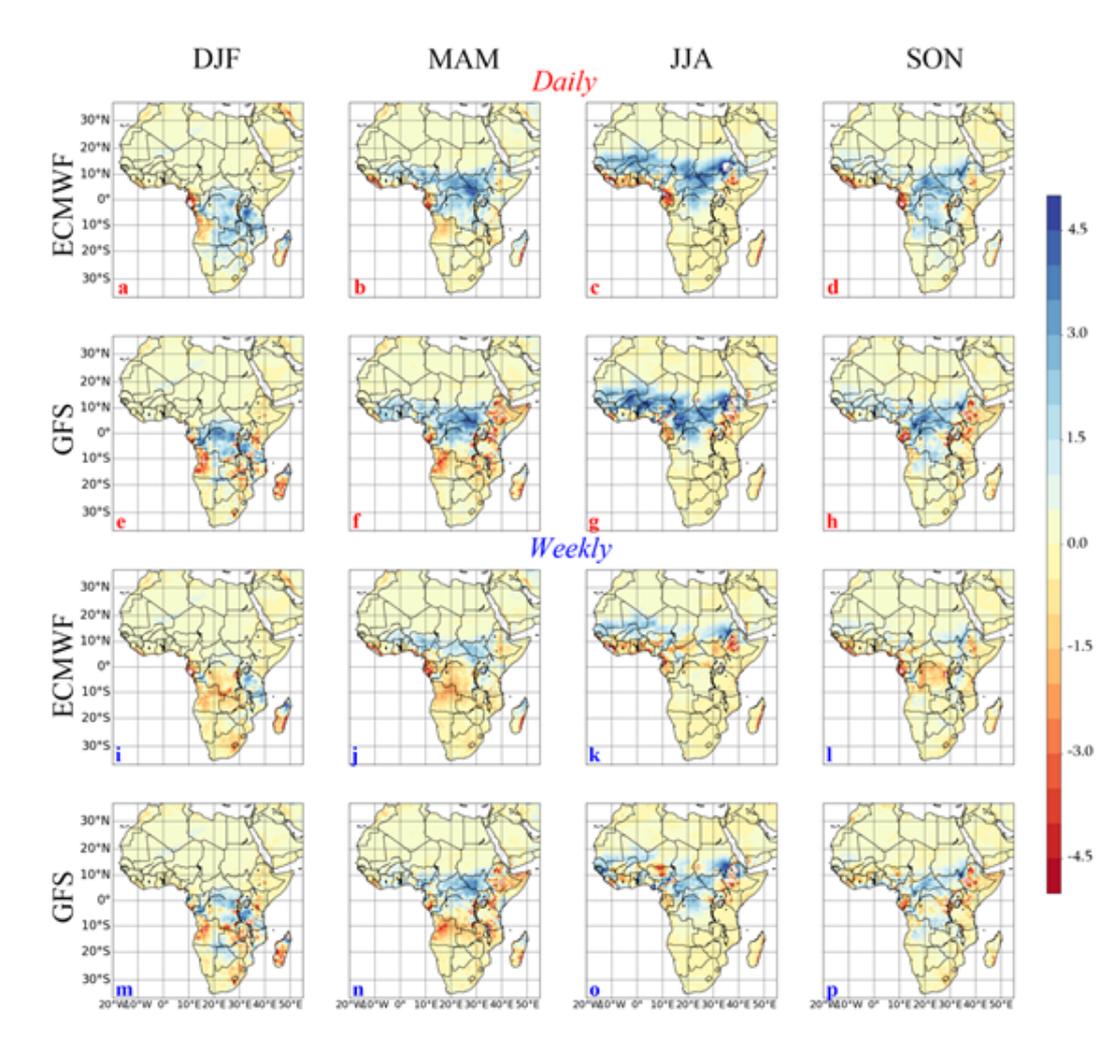

Figure S1. Spatial distribution of Bias (mean observation minus model; units: mm day$^{-1}$) between NWM forecasts (ECMWF and GFS) and ARC2 observations, at daily and weekly scales, in each season. Note: the NWP forecasts for daily rainfall prediction are at one day lead time.

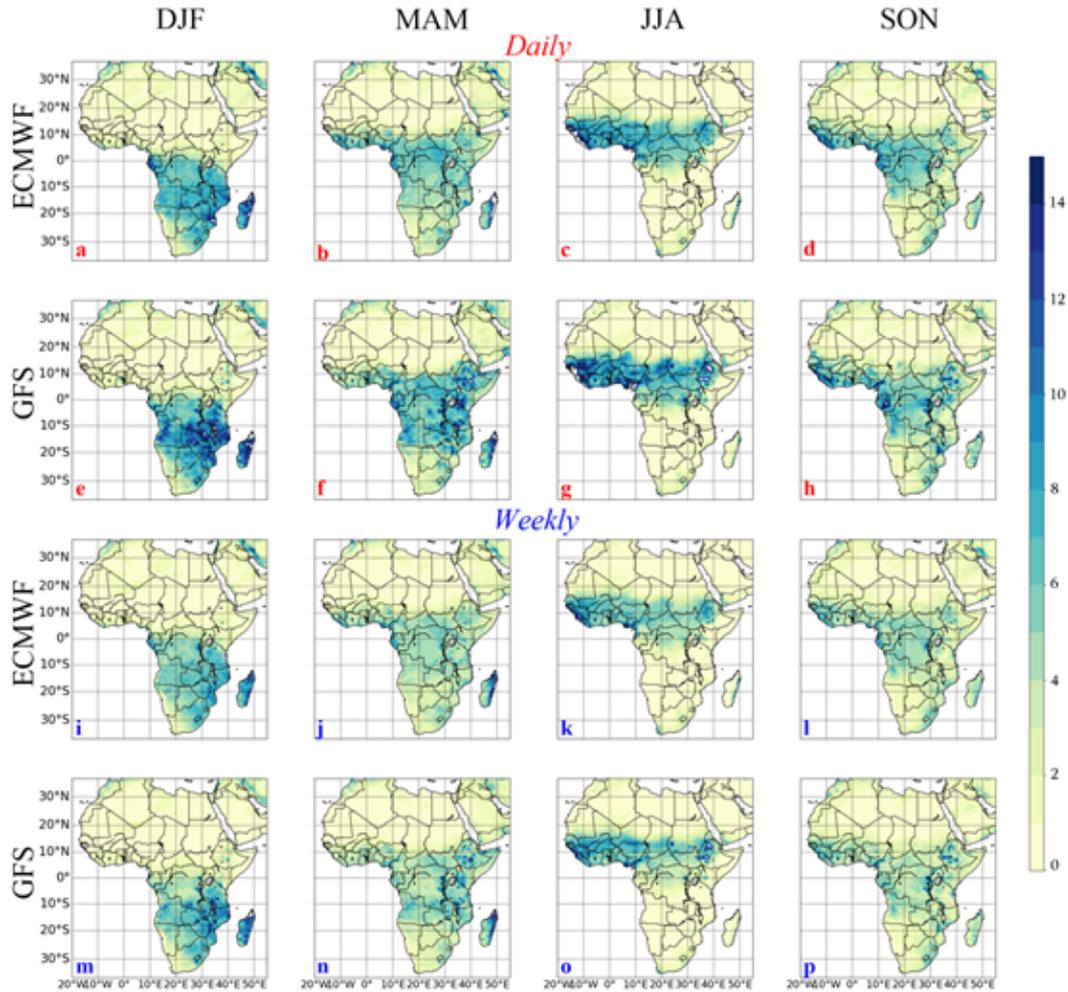

Figure S2. Spatial distribution of RMSE (units: mm day$^{-1}$) between NWM forecasts (ECMWF and GFS) and ARC2 observations, at daily and weekly scales, in each season. Note: the NWP forecasts for daily rainfall prediction are at one day lead time.

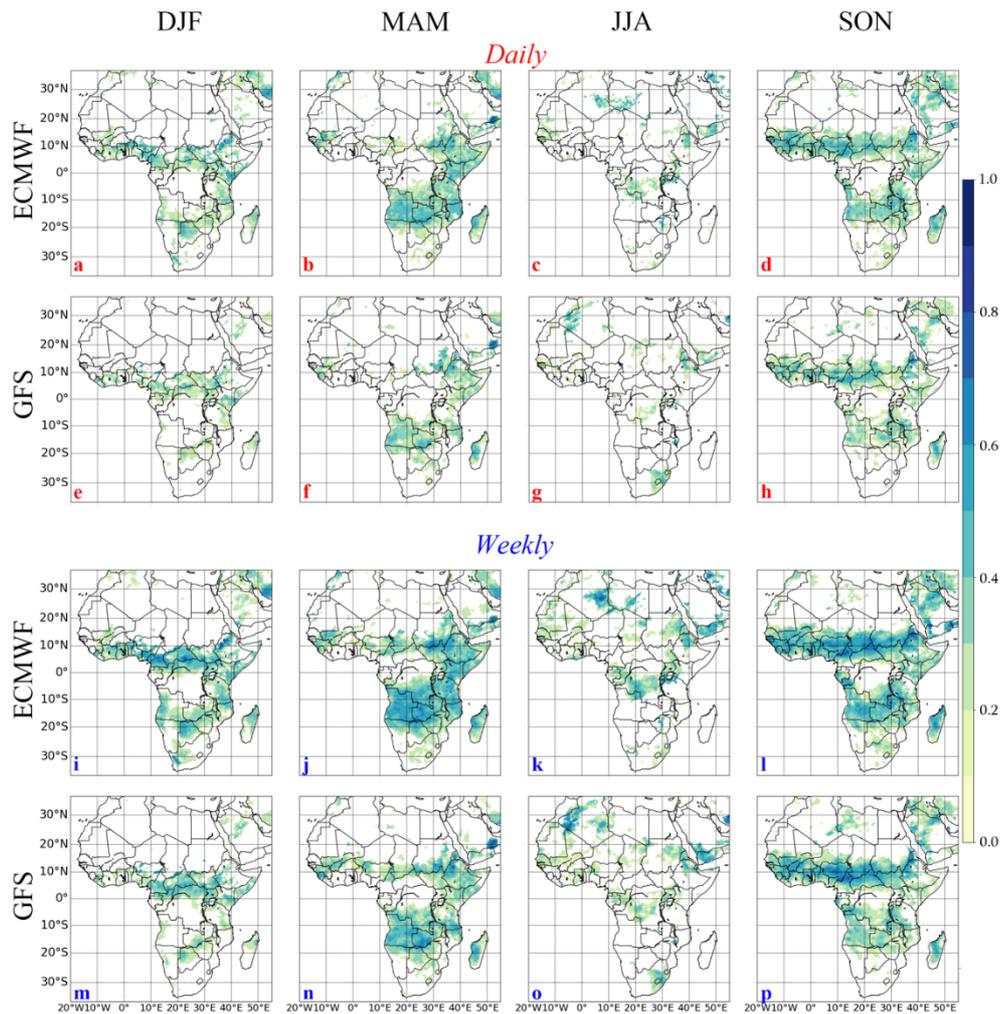

Figure S3. Spatial distributions of Correlation Coefficient (r) between NWM forecasts (ECMWF and GFS) and ARC2 observations, at daily and weekly scales, in each season. Note: the NWP forecasts are at one day lead time, and locations with effective days > 20 and P value < 0.01 (indicating positive skill) are shown in color.